\documentclass[conference]{IEEEtran}
\IEEEoverridecommandlockouts
\usepackage{cite}
\usepackage{amsmath,amssymb,amsfonts}
\usepackage{algorithmic}
\usepackage{graphicx}
\usepackage{textcomp}
\usepackage{xcolor}
\usepackage{textcomp}
\usepackage{booktabs} % For prettier tables
\usepackage{subfig}

\renewcommand{\baselinestretch}{0.98}
\def\BibTeX{{\rm B\kern-.05em{\sc i\kern-.025em b}\kern-.08em
    T\kern-.1667em\lower.7ex\hbox{E}\kern-.125emX}}
\begin{document}

\title{Robust Speed Control Methodology for Variable Speed Wind Turbines}

\author{\IEEEauthorblockN{Ammar Al-Jodah$^{*,1,2}$, Marwah Alwan$^2$}
\IEEEauthorblockA{\textit{$^1$Mechanical and Aerospace Engineering, Monash University, Australia}\\
\textit{$^2$Control and Systems Engineering, University of Technology, Baghdad, Iraq}\\
ammar.al-jodah@monash.edu}}
\maketitle

\begin{abstract}
Improving wind turbine efficiency is essential for reducing the costs of energy production. The highly nonlinear dynamics of the wind turbines and their uncertain operating conditions have posed many challenges for their control methods. In this work, a robust control strategy based on sliding mode and adaptive fuzzy disturbance observer is proposed for speed tracking in a variable speed wind turbine. First, the nonlinear mathematical model that describes the dynamics of the variable speed wind turbine is derived. This nonlinear model is then used to derive the control methodology and to find stability and robustness conditions. The control approach is designed to track the optimal wind speed that causes maximum energy extraction. The stability condition was verified using the Lyapunov stability theory. A simulation study was conducted to verify the method, and a comparative analysis was used to measure its effectiveness. The results showed a high tracking ability and robustness of the developed methodology. Moreover, higher power extraction was observed when compared to a classical control method.
\end{abstract}


\section{Introduction}
Renewable energy sources have received much attention in the recent years as they provide cleaner and cheaper energy as compared to fuel-based sources. The level of the carbon footprint that is produced from energy production plants has reached an alarming level around the globe. Carbon dioxide (CO2) is one of the major gases that emit from production plants, which has contributed significantly to global warming. The high level of CO2 pollution has adversely affected the ozone layer, the environment, and the quality of human lives in general. Moreover, it is estimated that the level of fossil fuel production will decline, and within a few generations from now, the world will exhaust all its reserve of fuel.  Therefore, there is an urgent need to explore renewable energy sources. Wind energy is one of these sources that can be exploited in many areas. It provides cheaper energy production with less required maintenance when compared to the solar-based solutions. Moreover, wind energy can provide power during the day and night, given the availability of the wind, unlike the solar-based energy. These impressive features have made this energy production sector one of the rapidly growing ones. And as the world energy demand is growing each year, it is expected for this sector to grow even more in the coming years. 

Variable speed wind turbines rotate with a variable speed that follows the speed of wind. Therefore, they achieve maximum utilization of the aerodynamic power. Most of the used wind turbines around the world are of the variable speed type due to less interruption in production and higher aerodynamic efficiency. However, the random and highly fluctuating wind speed has showed several challenges for their control. Controlling the wind turbine rotor under variable speed has become an active research topic in the recent years, as the wind industry started to adopt more advanced control techniques due to the deficiency of the classical control methods. 

Variable speed wind turbine has three regions of operation. In the first region, the wind turbine is just starting up. In the second region, the turbine has a normal operation mode, and it tries to maximize the capture of wind energy. The last region happens with high wind speed, and in this region, the turbine limits its rotor speed to protect its mechanical and electrical components. This work is focused on the second region of operation with the aim to optimize the wind energy extraction in this region through turbine’s rotor speed control.   

Many control approaches have been proposed to maximize the energy extraction from the wind turbine. However, the vast majority of these control approaches were linear and used a linearized version of the turbine model \cite{Boukhezzar2007}. These linear methods may succeed in attaining the desired level of power extraction and other set control requirements. Nevertheless, their performance may suffer greatly when subjected to parameter uncertainties, wind vibration, disturbances, and other unexpected external parameters \cite{Grimble1996}. Therefore the research in this area has been shifted toward exploring nonlinear control methods \cite{AlJodah2020a,AlJodah2020b}. A nonlinear control approach was proposed in \cite{Thomsen2007} to decouple disturbance from the speed control of a variable speed wind turbine. Feedback linearization was utilized in this process to transform the nonlinear system into a linear one. The wind fluctuations were reduced using this method, and a better performance was observed compared to LQG control. In \cite{Evangelista2010}, four nonlinear control strategies were tested for variable speed wind turbines. Sliding mode with variation law, super-twisting,  sub-optimal, and twisting were tested in simulation. It was found that all controllers were able to track the high-speed signal with no issues. Moreover, the super twisting was the easiest from the implementation perspective. Adaptive super twisting was developed in \cite{Evangelista2013} to maximize energy extraction. The strategy showed high robustness against uncertainties. An adaptive sliding mode control strategy was proposed in \cite{ Merabet2011}  for the speed tracking problem of variable speed turbines. The sliding surface is used in the update equation of the control gains, which resulted in improving the tracking results. In \cite{Narayana2009}, wind speed forecasting was used in variable speed turbines for maximum power point tracking. The prediction ability of this method improved its response time when compared to the traditional ones.  A fuzzy control system was developed in \cite{Zhang2006} for maximum power extraction. The Fuzzy controller provided fewer power fluctuations when compared with the classical PID. Another fuzzy controller was proposed in \cite{Yao2009} to maxmize power extraction through controlling turbine speed to follow a reference optimum speed. An active disturbance rejection method was proposed in \cite{Zhang2011} to maximize energy extraction via pitch angle control. The effectiveness of the method was proved in simulation with various wind profiles. A torque control method was proposed in \cite{Liao2009}, to control a variable speed wind turbine. The study investigated the utilization of the rotor side and the generator side speed signal for the torque control. It was found that the generator side provides better stability than the other one. A fuzzy control method was proposed in \cite{Boukhezzar2007} to have smooth output power. The method enhanced power regulation when compared with the PID control. Other control methods were proposed to solve control challenges in wind turbines such as second order sliding mode \cite{Zhang2021}, H$_{\infty}$\cite{FarajiNayeh2020},  switching sector sliding mode \cite{Berrada2020}, just to name a few. However, in these methods, the control action is selected to have a value that is higher than the maximum bound of the disturbance signal. This cause high load on the generator, and chattering in the signal for the discontinuous controllers. Therefor, this works reduces the demand on the control action to be higher than the disturbance estimation error instead of the disturbance itself.   

In this work, robust control methodology is proposed to make the rotor's speed of a variable speed wind turbine follow a trajectory that maximize power extraction. A PI-based sliding surface is used to have a high level of tracking performance. The stability analysis was conducted and verified using the Lyapunov stability criteria. Moreover, the necessary conditions for stability were found. A simulation study was conducted to verify the proposed strategy, and the results were compared with the classical PID controller. The results confirmed the high level of accuracy and robustness of the proposed approach.

\section{Modeling}
\begin{figure}[h!]
	\centering
	\includegraphics[width=0.8\linewidth]{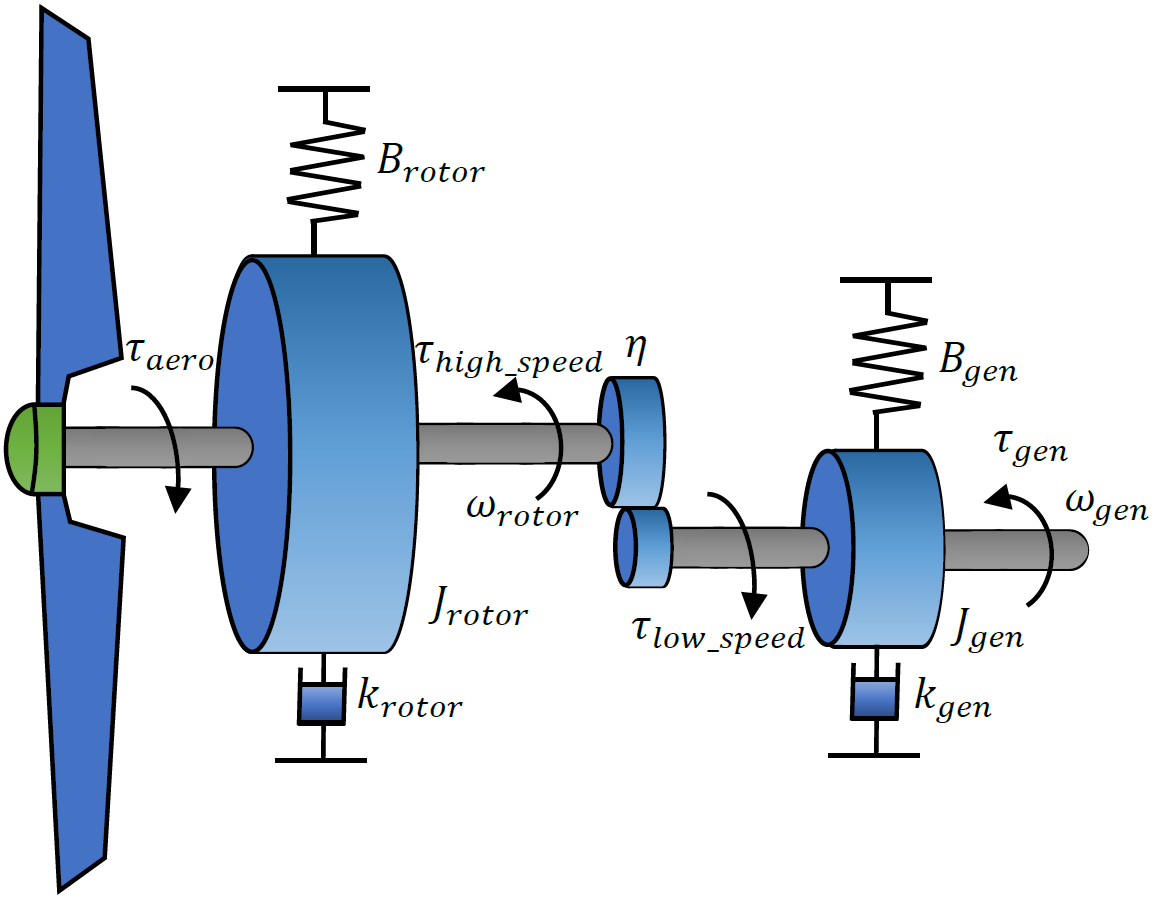}
	\caption{Wind turbine two mass model}
	\label{fig:WT_2mass}
	\vspace{-10pt}
\end{figure}

\begin{figure}[h!]
	\centering
	\includegraphics[width=\linewidth]{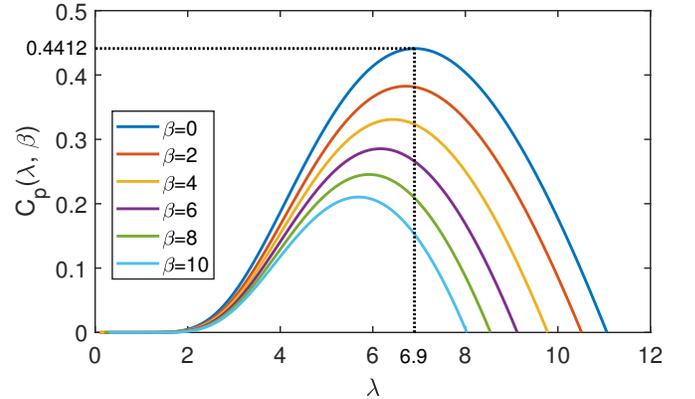}
	\caption{Wind turbine power coefficient curve.}
	\label{fig:cp_curve}
\end{figure}

The block diagram of the wind turbine is shown in Fig. \ref{fig:WT_2mass}. The variable speed wind turbine ia made of generator, aeroturbine, and a gearbox. The turbine rotor captures the aerodynamic power provided by \cite{Thomsen2007}
\begin{equation}
J_{aero} = \frac{1}{2} \rho \pi R^2 C_p(\lambda, \beta) v^{3}_{wind}
\end{equation}
where  $R$ is the rotor radius, $\beta$ is the blade pitch angle, $\rho$ is the air density, $v_{wind}$ wind speed, and $\lambda$ is the tip speed ratio and it is given by 
\begin{equation}\label{eq:lambda}
\lambda =  \omega_{rot} R/ v_{wind}
\end{equation}
where $\omega_{rot}$ is the rotor's speed. A nonlinear function of $\lambda$ and $\beta$ can describe the power coefficient $C_p$, and it is given by \cite{Slootweg2003}
\begin{equation}
C_p(\lambda, \beta) = \left( \mu_1 \ \varphi(\lambda, \beta) - \mu_2 \beta - \mu_3 \beta^x - \mu_4 \right)e^{-\mu_5 \varphi(\lambda, \beta)}
\end{equation}
\begin{equation}
\varphi(\lambda, \beta) = \frac{\beta^3 + 6 \times 10^{-6}\beta - 3\times 10^{-3} \lambda + 1}{-0.02 \beta^4 + \lambda \beta^3 -0.02 \beta + \lambda}
\end{equation}
where $\mu_1=110.23$, $\mu_2= 0.4234$, $\mu_3= 0.00146$, $\mu_4= 9.636$, $\mu_5= 18.4$, and $x= 2.14$. The maximum power coefficient is achieved when $beta=0$. This can be seen clearly from the 2D shape shown in Fig. \ref{fig:cp_curve}. The aerodynamic power is relative to the rotor torque by 
\begin{equation}
\tau_{aero} = \frac{P_{aero}} {\omega_{rot}}
\end{equation}
therefore
\begin{equation}
\tau_{aero} = \frac{1}{2 \omega_{rot}} \rho \pi R^2 C_p(\lambda, \beta) v^{3}_{wind}
\end{equation}
The rotor and generator dynamics equations are given by 
\begin{equation}\label{eq:Jr}
J_{rot}\dot{\omega}_{rot} = \tau_{aero} - \tau_{low\_speed} - B_{rot}\omega_{rot} - k_{rot} \theta_{rot}
\end{equation}
\begin{equation}
J_{gen}\dot{\omega}_{gen} = \tau_{high\_speed} - \tau_{elec} - B_{gen}\omega_{gen} - k_{gen} \theta_{gen}
\end{equation}
The gear box ratio is given by 
\begin{equation}
\eta = \frac{\tau_{low\_speed}}{\tau_{high\_speed}} = \frac{\omega_{gen}}{\omega_{rot}} = \frac{\theta_{gen}}{\theta_{rot}}
\end{equation}
Substituting the gear ratio in the generator equation gives 
\begin{multline}\label{eq:Jgen}
\eta^2 J_{gen}\dot{\omega}_{gen} = \tau_{low\_speed} - \eta \tau_{elec} \\ - \eta^2 B_{gen}\omega_{rot} - \eta^2 k_{gen} \theta_{rot}
\end{multline}
Adding Eq. (\ref{eq:Jgen}) and Eq. (\ref{eq:Jr}) results in 
\begin{equation}
J_{t}\omega_{rot} = \tau_{aero} - \tau_{gen} - B_{t}\omega_{rot}
\end{equation}
\begin{multline}
(J_{rot} + \eta^2 J_{gen})\dot{\omega}_{rot} = \tau_{aero} - \eta \tau_{elec} \\ - (B_{rot}+ \eta^2 B_{gen}) \omega_{rot} - (k_{rot} - \eta k_{gen} ) \theta_{rot}
\end{multline}
The model can be reduced to a single mass system, therefore the final dynamics equation is given by
\begin{equation}
J_{t}\dot{\omega}_{rot} = \tau_{aero} - \tau_{gen} - B_{t}\omega_{rot} - k_{t}\theta_{rot}
\end{equation}
where $J_{t} = J_{rot} + \eta^2 J_{gen}$, $J_{gen} = \eta \tau_{elec}$, $B_{t} = B_{rot} + \eta^2 B_{gen}$, and $k_{t} = k_{rot} + \eta k_{gen}$. The stiffness has negligible effect on the dynamics. Thus, by ignoring the stiffness and considering the uncertainty in parameters and external disturbances as a lumped disturbance $d$, the final dynamics equation is given by 
\begin{equation}
\dot{\omega}_{rot} = \frac{1}{J_{t}}\tau_{aero} - \frac{B_{t}}{J_{t}}\omega_{rot} - \frac{1}{J_{t}} \tau_{gen} - \frac{1}{J_{t}}d
\end{equation}
\section{Control method design}
It has been explained earlier that wind turbines operate in three regions. The current work is concerned with the second region, where the goal is to extract the highest amount of the wind energy. Therefore the control method objective is make the power coefficient $C_p$ at its peak. According to the $C_p$ curves given in Fig. \ref{fig:cp_curve}, it is desired to design a controller that keeps the $C_p$ at its highest possible value, i.e    
\begin{equation}
C_p(\lambda^{optimal}, \beta^{optimal}) = C^{optimal}_{p}
\end{equation}
To maximize $C_p$ it is necessary to keep $\lambda$, and $\beta$ at their optimal values. As indicated in Fig. \ref{fig:cp_curve}, the curve is maximized with $\beta^{optimal} = 0$, thus $\beta$ is set to zero in this work. The tip speed is the ratio of rotor to wind speed as was shown in Eq. (\ref{eq:lambda}). To keep it equivalent to its optimal value, it means the ratio should be kept constant. According to Fig. \ref{fig:cp_curve}, the optimal values for $C_p$ and $\lambda$ are given by $\lambda^{optimal}=6.9$, and $ C^{optimal}_{p} = 0.4412$. The wind speed is out of our control, thus that means the fluctuations in the wind speed should be tracked by the rotor to maintain $\lambda$ constant and equivalent to its optimal value. Therefore  
\begin{equation}
\omega^{optimal}_{rot} = \omega_{ref} = \lambda^{optimal} v_{wind}/R
\end{equation}

The control methodology will be designed to make the rotor speed follow $\omega^{optimal}_{rot}$, it will be refer to as $\omega_{ref}$ from now on. 
Defining the sliding variable as 
\begin{equation}
s = k_p\dot{e} + k_i e
\end{equation}
where $e = \omega_{rot} - \omega_{ref}$, differentiating $s$ to get 
\begin{equation}
\dot{s} = k_p(\dot{\omega}_{rot}-\dot{\omega}_{ref}) + k_i e
\end{equation}
substituting the rotor dynamics will result in 
\begin{multline}
\dot{s} = \frac{k_p}{J_{t}} \tau_{aero} - \frac{k_p B_{t}}{J_{t}}\omega_{rot} - \frac{k_p}{J_{t}}\tau_{gen} - k_p \dot{\omega}_{ref} \\ + \frac{k_p d}{J_{t}} + k_i e
\end{multline}
The stability is investigated by using the following Lyapunov function 
\begin{equation}
V_1= \frac{1}{2}s^2
\end{equation}
to have a stable system it is necessary to have a negative definite $\dot{V}_1$ \cite{AlJodah2020c,AlJodah2021}, thus   
\begin{multline}
\dot{V}_1 = s \dot{s} = s \left( \frac{k_p}{J_{t}} \tau_{aero} - \frac{k_p B_{t}}{J_{t}}\omega_{rot} - \frac{k_p}{J_{t}}\tau_{gen}\right. \\ \left.- k_p \dot{\omega}_{ref} + \frac{k_p d}{J_{t}} + k_i e\right)
\end{multline}
The control action $\tau_{gen}$ is designed to have two components as follows 
\begin{equation}
\tau_{gen} = u_{eq} + u_{sw} 
\end{equation}
The equivalent component is designed to maintain $s=0$, thus guarantee the high accuracy of the motion tracking. Therefore it is designed as 
\begin{equation}
u_{eq} = \tau_{aero} - B_{t}\omega_{rot} - J_{t} \dot{\omega}_{ref}  + \frac{J_{t}k_i}{k_p}e + \hat{d}
\end{equation} 
where $\hat{d}$ is the estimated disturbance. The switching control action is designed to bring the system states to the sliding variable and to reject the system's uncertainties \cite{Ghafarian2020}. It is designed as 
\begin{equation}
u_{sw} = \frac{J_{t}}{k_p}(k_1s+k_2sgn(s))
\end{equation}   
substituting $\tau_{gen}$ in $\dot{V}_1$ to check the stability condition as follows 
\begin{align}
\dot{V}_1 &= s\left(-k_1 s - k_s sgn(s) + \frac{k_p}{J_{t}}\tilde{d}\right)\\
&= -k_1 s^2 - \left(k_2 |s| -   \frac{k_p}{J_{t}}\tilde{d} |s|\right)
\end{align} 
therefore 
\begin{align}
-k_1 s^2 - \frac{k_p}{J_{t}} |s|\left(\frac{J_{t}k_2}{k_p} - |\tilde{d}|\right) & \le 0 \\
\end{align}
The stability condition is given by 
\begin{equation}\label{eq:stab_cond_v1}
\frac{J_{t}k_2}{k_p} > |\tilde{d}|
\end{equation}
This indicates that the system will be stable as long as the disturbance estimation is bounded and the control gain is higher than this bound. In this work, Adaptive Fuzzy Disturbance Observer (AFDO) is designed to estimate the disturbance and hence reduce the demand on the control action to be only higher than the bound of disturbance estimation error rather than the disturbance itself, as evident from the stability condition above. 
According to fuzzy basis function, the disturbance can be estimated by 
\begin{equation}\label{eq:fuzzy_output_re}
\hat{d}(\bar{\omega}|\hat{\theta})=\hat{\theta}^T\psi(\bar{\omega})
\end{equation}
where $\bar{\omega}=[{\omega}_{1},\ {\omega}_{2}]^T=[{\omega}_{rot},\ \dot{\omega}_{rot}]^T$, $\hat{\theta}$  is an adaptive parameter, $\psi(\bar{\omega})$ is the fuzzy basis function, both of dimension $p$, where $p=m^2$, $m$ is the number of membership functions. The fuzzy basis function can be defined as 
\begin{equation}\label{eq:fuzzy_basis}
\psi_{l_1,l_2}(\bar{\omega})=\frac{\prod_{i=1}^{2} \mu_{A_{i}^{l_i}} (\bar{\omega}_{i})} {\sum_{l_1=1}^{m}\sum_{l_2=1}^{m} \left (\prod_{i=1}^{2} \mu_{A_{i}^{l_i}} (\bar{\omega}_{i})\right )}
\end{equation}
where $l_1,\ l_2 = 1 \to m$. The AFDO design is based on the following assumption.

\textbf{\textit{Assumption \cite{AlJodah2020}}}: the optimal  adjustable parameter $\theta^*$ will result in a bounded disturbance estimation that is given by $\displaystyle\sup_{\omega \in M_{\bar{\omega}}}\left | d(\omega)-\hat{d}(\omega|\theta^*)\right|< \varepsilon$, where $M_{\bar{\omega}}$ is a compact set, and $\varepsilon$ is a non-negative scalar.\newline
The AFDO is defined by 
\begin{equation}\label{eq:z_dot}
\dot{z}=-\sigma z + p(\bar{\omega},u,\hat{\theta})
\end{equation}
where $z$ is the observer state, $p(\bar{\omega},u,\hat{\theta})=\sigma \omega_2 + \frac{1}{J_t}\tau_{aero} - \frac{B_t}{J_t}\omega_1-\frac{1}{J_t}\tau_{gen}-\frac{1}{J_t}\hat{d}(\bar{\omega}|\hat{\theta})$, and $\sigma$ is a positive constant. The disturbance estimation error is given by
\begin{equation}\label{eq:FDO_dist_observ_err}
\zeta=x_2-z
\end{equation}
Thus, Eq. (\ref{eq:z_dot}) can be rewritten as
\begin{equation}\label{eq:z_dot_2}
\dot{z}=\sigma\zeta + \frac{1}{J_t}\tau_{aero} - \frac{B_t}{J_t}\omega_1-\frac{1}{J_t}\tau_{gen}-\frac{1}{J_t}\hat{d}(\bar{\omega}|\hat{\theta})
\end{equation}
The adjustable $\hat{\theta}$ is adapted to reconstruct $\hat{d}(\bar{\omega}|\hat{\theta})$ which in turn  estimate $d(\bar{\omega})$. Therefore, the error dynamics should be established as follows   
\begin{equation}\label{eq:FDO_error_dynamics}
\dot{\zeta}=-\sigma\zeta + \frac{1}{J_t} \left(\hat{d}(\bar{\omega}|\hat{\theta}) - d(\bar{\omega}) \right )
\end{equation}
According to the previously stated assumption and the capacity of fuzzy system to act as a universal 
approximator, $d(\bar{\omega})$ can be defined as 
\begin{equation}\label{eq:disturbance_fdo}
d(\bar{\omega})= {\theta^*}^T\psi(\bar{\omega}) + \varepsilon
\end{equation}
where $*$ stands for the optimal value, and  $\varepsilon$ is the estimation error. Thus, the following error dynamics can be found    
\begin{align}\label{eq:FDO_error_dynamics2}
\dot{\zeta} & = -\sigma\zeta+ \frac{1}{J_t}\hat{\theta}^T\psi(\bar{\omega}) - \frac{1}{J_t}{\theta^*}^T\psi(\bar{\omega})  -\frac{1}{J_t}\varepsilon\nonumber\\
& = -\sigma\zeta+ \frac{1}{J_t}\tilde{{\theta}}^T\psi(\bar{\omega}) + \epsilon
\end{align}
where $\epsilon = -(1/J_t)\varepsilon$, and $\tilde{\theta}^T=\hat{\theta}^T - {\theta^*}^T$. The adaptive law of $\hat{\theta}$ can be found using the following Lyapunov function
\begin{equation}\label{eq:FDO_Lyapunov}
V_2= V_1 + \frac{1}{2}\zeta^2 + \frac{1}{2\gamma}\tilde{\theta}^T\tilde{\theta}
\end{equation}
A stable feedback system can be guaranteed when $\dot{V}_2$ is negative semi-definite. Differentiating $V_2$ provides
\begin{align}\label{eq:FDO_V_dot}
\dot{V}_2 & = \dot{V}_1 + \zeta \dot{\zeta} +\frac{1}{\gamma}\tilde{\theta}^T \dot{\tilde{\theta}}\nonumber \\
& =\dot{V}_1 -\sigma\zeta^2 + \frac{1}{J_t}\zeta\tilde{\theta}^T\psi(\bar{\omega}) + \zeta\epsilon +\frac{1}{\gamma}\tilde{\theta}^T \dot{\tilde{\theta}}\nonumber \\
& = \dot{V}_1 -\sigma\zeta^2 + \tilde{\theta}^T\left( \frac{1}{J_t}\zeta\psi(\bar{\omega})+ \frac{1}{\gamma}\dot{\tilde{\theta}} \right) +  \zeta\epsilon
\end{align}
Defining $\dot{\tilde{\theta}}$ as
\begin{equation}\label{eq:th_tild_adapt_law}
\dot{\tilde{\theta}}= -\frac{\gamma}{J_t}\zeta\psi(\bar{\omega}) = -\bar{\gamma}\zeta\psi(\bar{\omega})
\end{equation}
where $\bar{\gamma}=\gamma/J_t$, and by using this definition, the following adaptive law can be established 
\begin{equation}\label{eq:th_adapt_law}
\dot{\hat{\theta}}=\bar{\gamma}\zeta\psi(\bar{\omega})
\end{equation}
Substituting Eq. (\ref{eq:th_tild_adapt_law}) in Eq. (\ref{eq:FDO_V_dot}) will yield
\begin{align}\label{eq:FDO_V_dot_final}
\dot{V}_2 &= \dot{V}_1 -\sigma\zeta^2 +\zeta\epsilon\nonumber\\
& = \dot{V}_1 - \frac{\sigma}{2}\zeta^2 + \frac{1}{2\sigma}\epsilon^2 - \left( \sqrt{\frac{\sigma}{2}}\zeta - \sqrt{\frac{1}{2\sigma}}\epsilon \right)^2\\
& \leq \dot{V}_1  - \frac{\sigma}{2}\zeta^2 + \frac{1}{2\sigma}\epsilon^2\nonumber\\
& \leq - \frac{k_p}{J_{t}} |s|\left(\frac{J_{t}k_2}{k_p} - |\tilde{d}|\right)  - \frac{\sigma}{2} \left( \zeta^2 - \frac{\epsilon^2}{\sigma^2}\right)
\end{align}
For $\dot{V}$ to be negative definite the following condition for $\zeta$ should hold: $ \zeta^2 \geq \epsilon^2/\sigma^2$, along with the condition given in Eq. (\ref{eq:stab_cond_v1}). 

It is worth noting here is that the designed approach needs the differentiation of the reference rotor speed. The reference signal is related to the wind speed, and that has high fluctuations. A low pass filter was utilized with the differentiator to reduce the noise amplification during the differentiation process. The switching action is depending on the nonlinear signum function, which, when used, introduces chattering in the control signal. Therefore, it has been replaced with a more smooth function $tanh$.
\begin{figure}[h!]
	\centering
	\includegraphics[width=\linewidth]{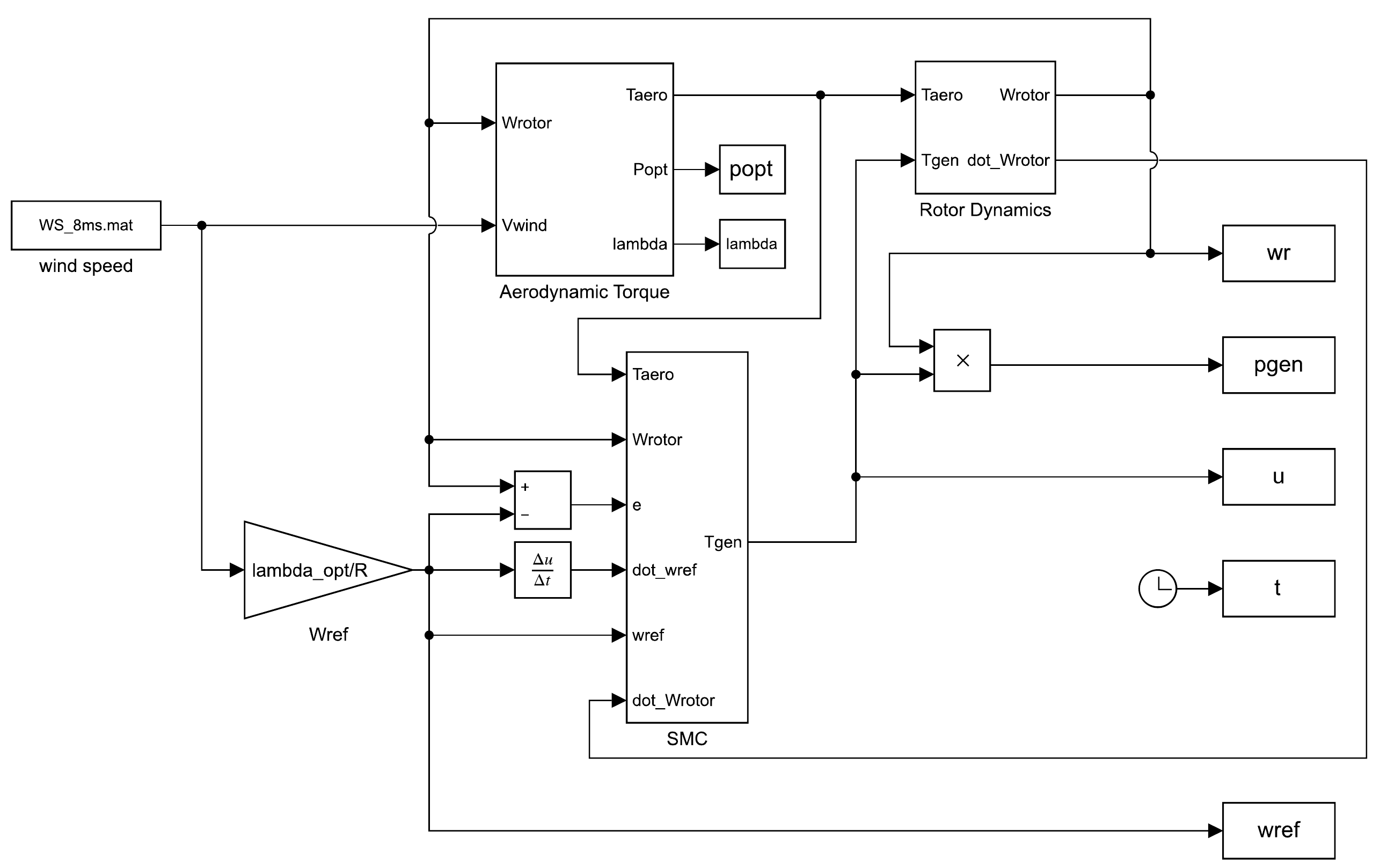}
	\caption{Feedback system block diagram.}
	\label{fig:sys_bd}
	\vspace{-10pt}
\end{figure}
\begin{figure}[h!]
	\centering
	\includegraphics[width=\linewidth]{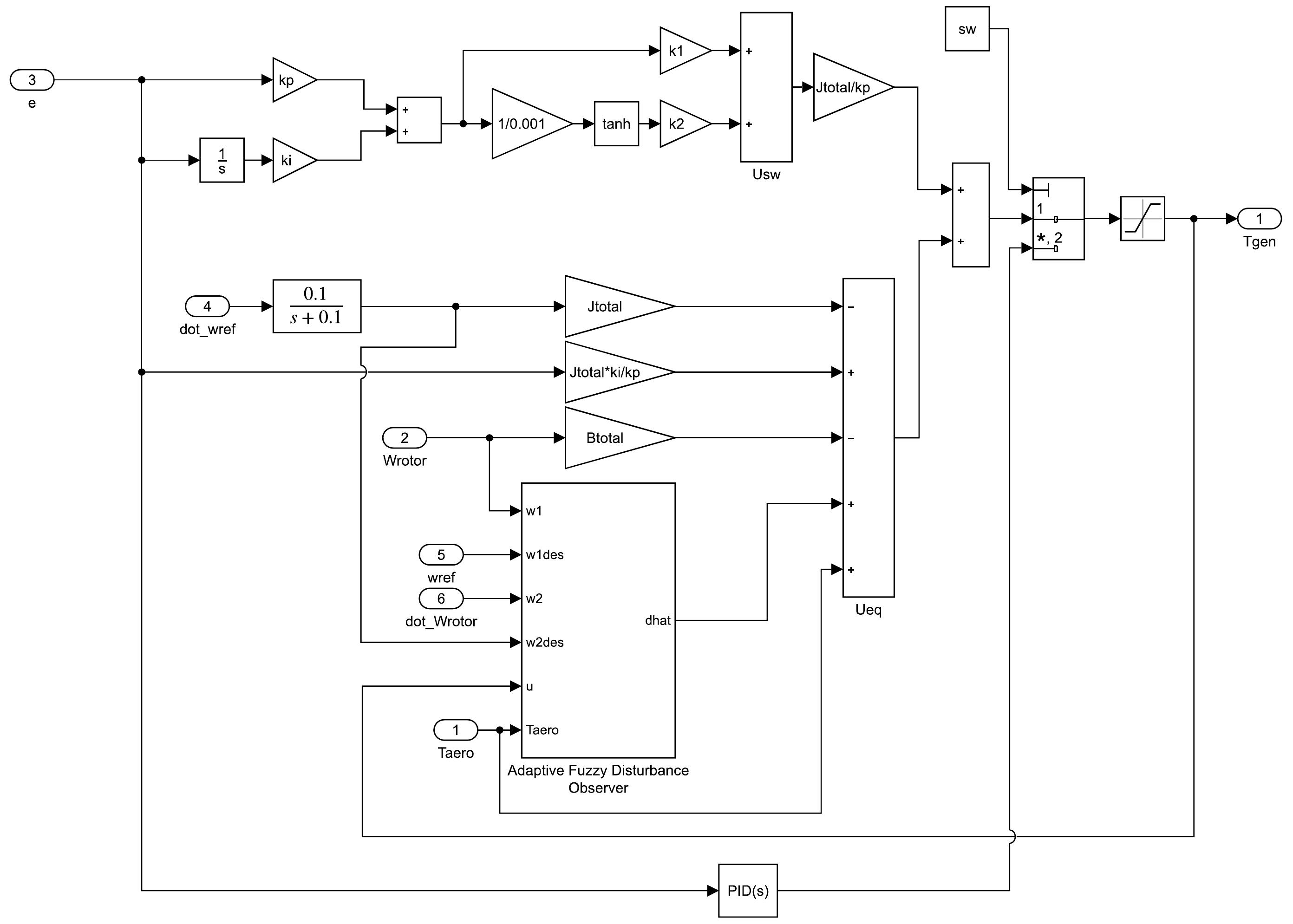}
	\caption{Sliding mode controller block diagram.}
	\label{fig:smc_bd}
\end{figure}
\section{Results}
\begin{figure}[!h]
	\centering
	\includegraphics[width=\linewidth]{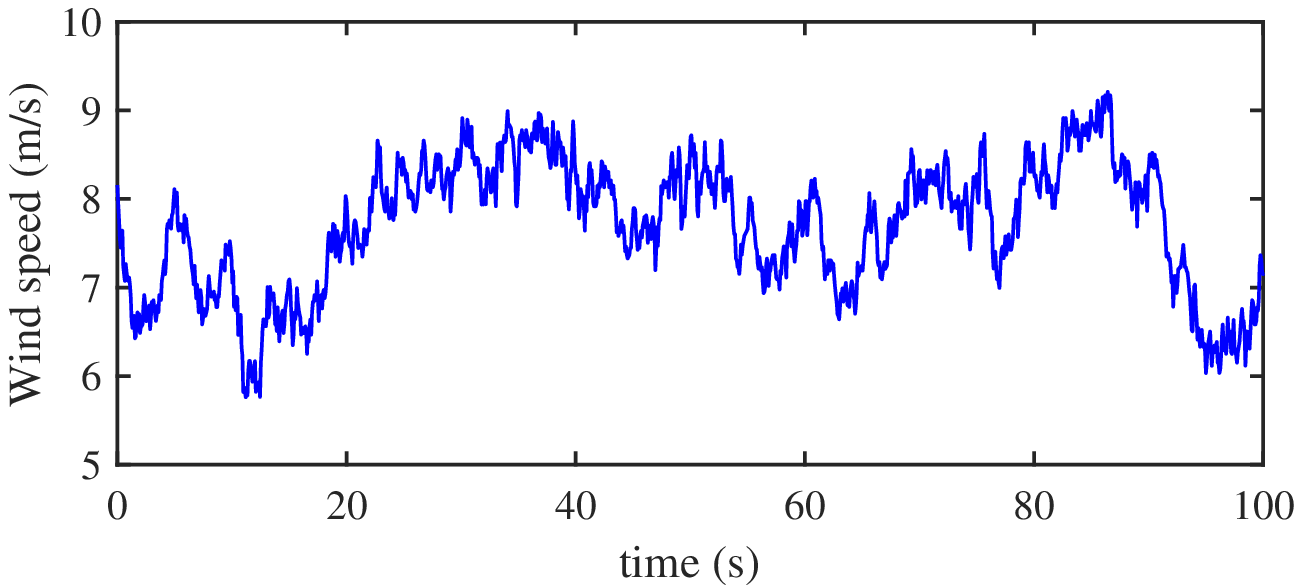}
	\caption{8 m/s mean wind speed profile}
	\label{fig:wind8}
	\vspace{-10pt}
\end{figure}

\begin{figure}[!h]
	\centering
	\includegraphics[width=\linewidth]{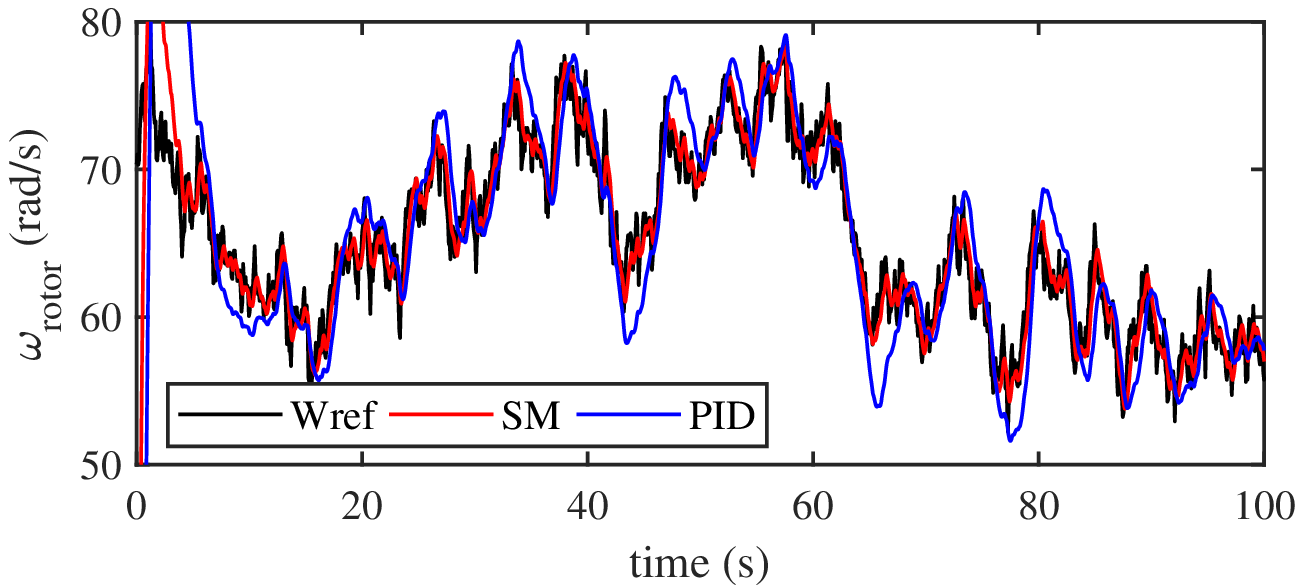}
	\caption{rotor speed tracking}
	\label{fig:wr}
	\vspace{-10pt}
\end{figure}

\begin{figure}[!h]
	\centering
	\includegraphics[width=\linewidth]{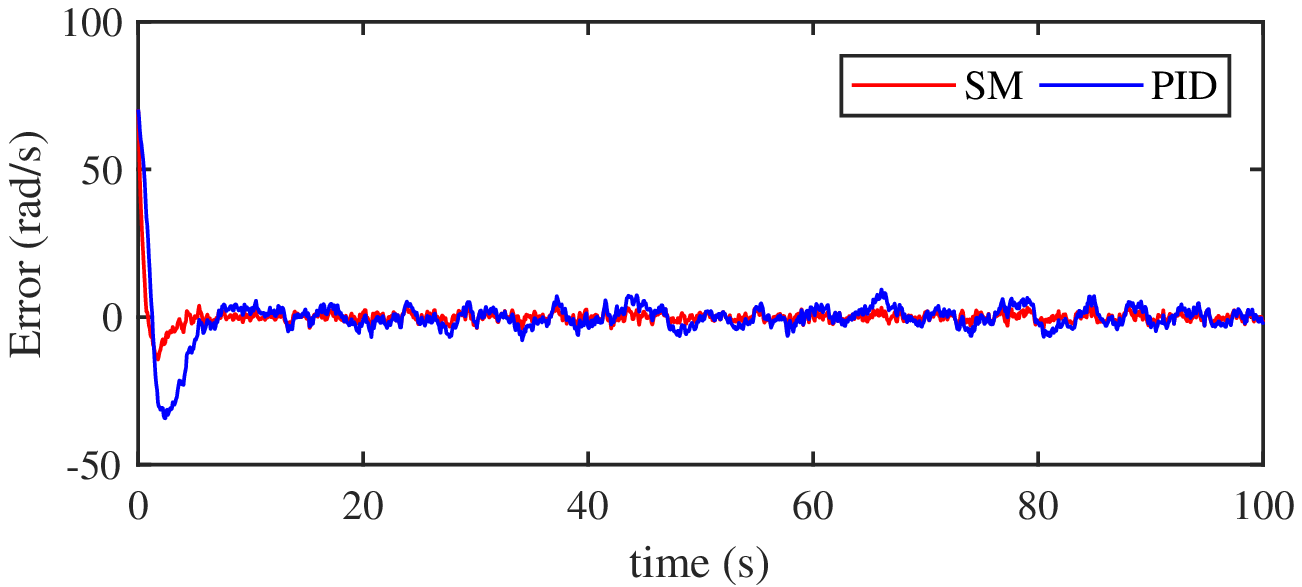}
	\caption{rotor speed tracking error}
	\label{fig:err}
	\vspace{-10pt}
\end{figure}

Simulink was used to simulate the wind turbine system and its control approach. The overall feedback system as was designed in Simulink is shown in Fig. \ref{fig:sys_bd}. The control approach is designed as in Fig. \ref{fig:smc_bd}. The signum function was replaced by $tanh(s/0.001)$ to reduce chattering \cite{Ghafarian2020b,AlJodah2018,AlJodah2018b,Das2020,Ghafarian2020c,humaidi2018}.  The $\dot{\omega}_{ref}$ was produced by differentiating the reference speed and filtering the results to reduce the noise. To compare the results with a PID controller a switching block was used to switch between strategies. The model parameters are presented in Table \ref{tab:param}.
\begin{table}[h!]
	\centering
	\caption{Model parameters.}
	\label{tab:param}
	\begin{tabular}{cccc}
		\toprule 
		\textbf{Parameter}  & \textbf{Value} & \textbf{Parameter}  & \textbf{Value} \\
		\midrule 
		$\rho$ & 1.29 &  $Bt$ & 0\\
		$Jt$ &  1.5 & $C^{optimal}_{p}$ & 0.48\\
		$R$ & 1.26 & $\lambda^{optimal}$ & 6.9\\
		\bottomrule
	\end{tabular}
	\vspace{-10pt}
\end{table}
A wind profile with a speed mean value of 8m/s  was used in the simulation, as shown in Figs. \ref{fig:wind8}. The sliding mode (SM) method was able to track the reference speed with high accuracy when compared with the PID controller as shown in Figs \ref{fig:wr}, and \ref{fig:err}.The mean square error for the PID controller was 15.6675, while for SM was 4.3684, therefore the SM method has improved the tracking results by 72\%. It was noticed that the PID controller has higher overshoot than the SM. 

\begin{figure}[!h]
	\centering
	\includegraphics[width=\linewidth]{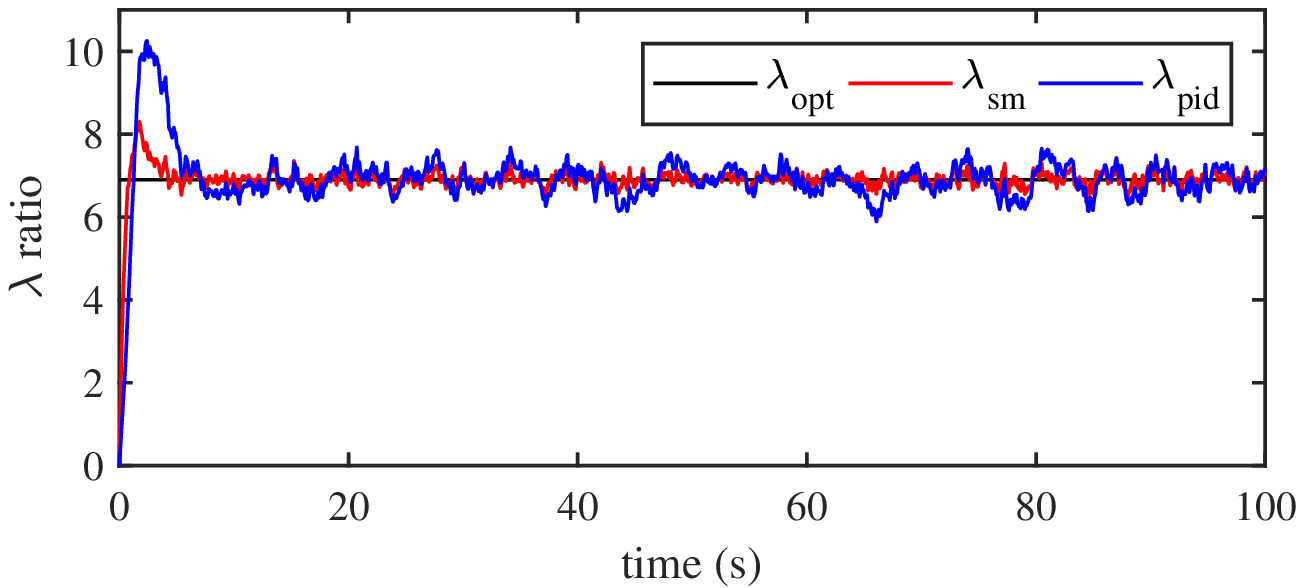}
	\caption{tip speed $\lambda$}
	\label{fig:lam}
	\vspace{-10pt}
\end{figure}

The tip speed tracking of the optimal value is shown in Fig. \ref{fig:lam}. The mean square error of $\lambda$ tracking for the PID controller was 0.4667, while for SM was 0.1194, therefore the SM method has improved the tracking results by 74\%. Again PID showed high overshoot.  

\begin{figure}[!h]
	\centering
	\includegraphics[width=\linewidth]{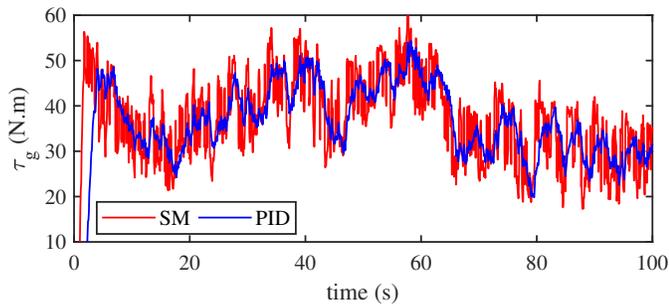}
	\caption{generator torque}
	\label{fig:T}
	\vspace{-10pt}
\end{figure}

\begin{figure}[!h]
	\centering
	\includegraphics[width=\linewidth]{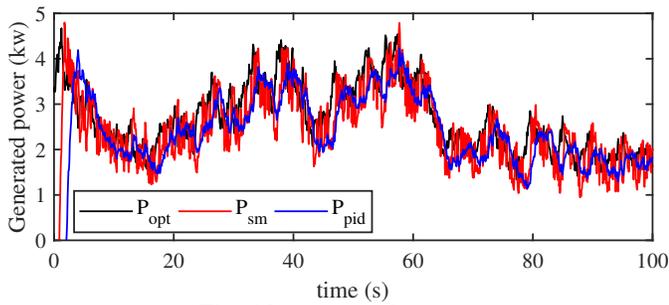}
	\caption{generated power}
	\label{fig:P}
	\vspace{-10pt}
\end{figure}

The transient response of the control system has seen high spikes for the control action (the generated torque), Fig \ref{fig:T}. This high initial value of the torque is due to the high magnitude of the error, external disturbance, and uncertainty in the parameters. The control action becomes high to overcome all these uncertainties. This high peak of the control signal soon reduces in intensity after a few seconds, and the curve follows the changes in the wind speed. Since the wind speed is not regular and randomly changes its shape, the control signal follows the same behavior. And it is seen as fluctuating or oscillating with the wind changes. The standard deviation of the torque signal for the PID was  7.7306, where for SM was 10.4137. Form these numbers, it can be seen that the SM control action has a higher variation than the PID controller. This is expected as SM needs to change quickly to follow the speed fluctuations. To measure the energy capture efficiency, the following equation is adopted
\begin{equation}
E = \frac{ \int_0^t P_{gen}(\tau) d\tau}{\int_0^t P^{optimal}_{aero}(\tau) d\tau} =  \frac{\int_0^t \tau_{gen}\omega_{gen}}{\int_0^t \frac{1}{2}\rho \pi R^2 C^{optimal}_p v^{3}_{wind}}
\end{equation}
The energy generation is shown in Fig. \ref{fig:P}. The energy capture efficiency for the PID controller was 88.47\%, while for SM was 89.79\%, therefore the SM method has improved the energy production by 1.32\%.
\section{Conclusion}
In this study, a robust control methodology was designed to optimize the energy extraction in a variable speed wind turbine. The nonlinear model of the wind turbine was developed, and the aerodynamics of the rotor was described as a nonlinear model that depends on the tip and rotor speed. The optimal tip speed and power coefficient were found from the power coefficient curves. These values than later used in the development of the control method to obtain maximum power production. Since the wind speed is out of the design method's control, the rotor speed was used to maintain the tip speed ratio. The sliding mode control method based on an proportional-integral sliding variable and adaptive fuzzy disturbance observer was utilized to maximize the rotor speed tracking accuracy. A simulation study was performed  using Simulink, and the results indicated the high energy generation and tracking ability of the robust control method compared with a PID control. Future studies can be focused on reducing the oscillation in the torque signal. 
\bibliographystyle{ieeetr}

\end{document}